# Enhanced Anomaly Detection in IoMT Networks using Ensemble AI Models on the CICIoMT2024 Dataset

**Prathamesh Bhairu Chandekar**
D Y Patil International University
Maharashtra, India
20220802022@dypiu.ac.in

**Mansi Swarupa Mehta**
D Y Patil International University
Maharashtra, India
20220802045@dypiu.ac.in

**Dr. Swet Chandan**
D Y Patil International University
Maharashtra, India
swet.chandan@dypiu.ac.in

**Abstract:**

The rapid proliferation of Internet of Medical Things (IoMT) devices in healthcare has introduced unique cybersecurity challenges, primarily due to the diverse communication protocols and critical nature of these devices. This research aims to develop an advanced, real-time anomaly detection framework tailored for IoMT network traffic, leveraging AI/ML models and the CICIoMT2024 dataset. By integrating multi-protocol (MQTT, WiFi), attack-specific (DoS, DDoS), time-series (active/idle states), and device-specific (Bluetooth) data, our study captures a comprehensive range of IoMT interactions. As part of our data analysis, various machine learning techniques are employed which include an ensemble model using XGBoost for improved performance against specific attack types, sequential models comprised of LSTM and CNN-LSTM that leverage time dependencies, and unsupervised models such as Autoencoders and Isolation Forest that are good in general anomaly detection. The results of the experiment prove that augmenting any neural network with an ensemble model lowers false positive rates significantly and reduces detections while offering an enhanced model applicable to IoMT. This framework offers critical concepts of systems and techniques for real-time healthcare anomaly detection and thus emphasizes the urgency for customized IoMT security measures.

## 1 Introduction

IoMT refers to specific technologies that leverage the connective power of the Internet to better healthcare services and treatment. As examples, wearable sensors, health monitors or diagnostic devices can transmit patient information constantly capturing, sending and providing analytics on the state of the health of a patient. Thanks to this ability to connect and share data at any time, healthcare management services can tailor their services to individual persons, to focus on prevention, and even to deliver therapy at a distance. A steady increase in the usage of IoMT in clinical practice and in home care settings has introduced an order of emphasis on these systems in the management of persistent diseases, enhancement of care outcomes, and trimming down of costs of healthcare services.

### 1.1 Challenges in IoMT Security

IoMT has brought several advantages in the healthcare sector but comes with a lot of risks regarding security and privacy. IoMT devices generally work with minimal processing power which makes the integration of most conventional security systems difficult. Such devices are susceptible to several cyber threats including DDoS, DoS, spoofing, and reconnaissance attacks that threaten to hinder healthcare service delivery and access to sensitive information about patients. Because such attacks may threaten the safety of patients and the integrity of their records, there is greater interest in the development of suitable anomaly detection systems for deployment in IoMT environments.

### 1.2 The Role of Machine Learning in IoMT Anomaly Detection

Machine Learning(ML) has been significant in improving IoMT security through adaptive anomaly detection techniques that handle the high dimensionality and heterogeneity of IoMT data. In contrast to the conventional rule-based systems, cyber–ML Mechanisms can process complex aspects of data and learn issues within the data which corresponds to deviations suggesting some forms of attacks. More complex ML models including Autoencoders, Isolation Forests, Long Short-Term Memory LSTM, and Convolutional Neural Network LSTM (CNN-LSTM) models perform better in identifying anomalies events from different types of data in IoMT. These models allow the requirements for real time responses to security threats as well as the ability to perform adaptive threat response making them appropriate for healthcare domain.

### 1.3 Need for Comprehensive IoMT Datasets

The efficacy of ML-based anomaly detection depends heavily on the availability of high-quality, comprehensive datasets. However, current datasets often lack diversity in terms of device types, protocols, and attack scenarios, limiting the generalizability of detection models. There is a pressing need for datasets that reflect the unique characteristics of IoMT traffic and attack profiles to support the development of robust, generalizable security solutions.

### 1.4 Objective and Contributions

This research leverages the CICIoMT2024 dataset, a unique, benchmark dataset tailored for IoMT anomaly detection. Our study aims to:

- Develop and evaluate advanced ML models for anomaly detection in IoMT networks.

- Apply models across various data segments within CICIoMT2024, including multi-protocol data, attack-specific data, time-series data, and device-specific profiles.

- Demonstrate the advantages of an ensemble approach that combines the strengths of individual models for robust and accurate anomaly detection.

Through these contributions, this research provides a foundation for secure IoMT deployments in healthcare, enabling more resilient and reliable patient care solutions.

## 2 Literature Review

The rapid adoption of IoMT devices in healthcare has intensified research into IoMT security, particularly in anomaly detection. Traditional security measures like firewalls and encryption are inadequate for IoMT environments due to the resource-constrained nature of IoMT devices and the complexity of healthcare data. The application of ML for anomaly detection in IoMT has received considerable attention, and models have been built emphasizing the supervised, semi-supervised, and unsupervised approaches. Nevertheless, these approaches usually fail due to the diversity of IoMT traffic, complexity of the attack patterns, and needs of real-time analysis and monitoring.

### 2.1 Machine Learning Models for Anomaly Detection in IoMT

Several ML models have already been examined in relation to the security of IoMT systems. In this regard, for example, Autoencoders and Isolation Forests have performed well in unsupervised anomaly detection, where there is limited availability of labelled data. LSTM networks and CNN-LSTM hybrids are used in most sequential analysis due to their ability to capture time dependencies in IoMT data. Similarly, ensemble methods, which combine multiple ML models to improve detection accuracy while minimizing false positives by exploiting the best features of each model, are also effective.

- **Autoencoders:** Autoencoders are Artificial Neural Networks trained in an unsupervised manner with the purpose of learning particular data representations through the minimization of reconstruction errors. In IoMT systems such networks may serve the purpose of detecting anomalies due to instance of high reconstruction error.

- **Isolation Forests:** Isolation forests are an ensemble method that separates anomalies by dividing their population recursively, which is very useful for outlier detection in a large and high dimensional dataset.

- **LSTM and CNN-LSTM Networks:** These models are ideal for sequential data because they learn the time dependencies within IoMT data to detect changes in the behaviour pattern.

### 2.2 Challenges with Existing Datasets

Many publicly available IoMT datasets have different shortcomings in terms of information and research. Datasets available lack the diversity of protocol types, device behaviours, and attack scenarios making it hard to build models that can be generalized and applied to real-world IoMT networks. The CICIoMT2024 dataset addresses this issue by including multiple protocols (Wi-Fi, MQTT, TCP/IP) and various attack types (DoS, DDoS) while also getting active and idle states per device and time series information capturing different device profiles. This dataset therefore allows a more extensive model validation process as well as development of models which in turn enhance the effectiveness of IoMT anomaly detection solutions.

### 2.3 Research Gap and Proposed Approach

Advancements have been made towards works in ML-based anomaly detection but there is still scope of improvement with respect to developing specific tools to IoMT networks which include the different types of devices, protocols, and real time aspects. The limitations are addressed in this work by implementing a multi-model approach on the CICIoMT2024 dataset consisting of Autoencoders, Isolation Forests, LSTM, CNN-LSTM hybrids and ensemble stacking. This augmenting of our research provides an original contribution to the subject of IoMT devices by systematically evaluating these models in different IoMT models which could serve as a reference point for the developers in IoMT security.

## 3 Dataset

CICIoMT2024 which was created by the Canadian Institute for Cybersecurity is one of a kind dataset that has been created for the purpose of IoMT anomaly detection. This dataset overcomes some inadequacies present in other available resources by covering various types of protocols, types of device and types of attacks. It consists of a total of four main parts including Multi-Protocol Data, Attack Specific Data, Time-series Data and Device Specific Profiles. The scope of each segment is such as to facilitate the assessment of ML models pertaining to selected IoMT security issues.

### 3.1 Dataset Structure and Key Characteristics

- **Multi-Protocol Data (Wi-Fi, MQTT):** This segment also encompasses benign and attack traffic from several protocols, which is closer to the reality of the IoMT environments. The presence of such data within the DDoS and port scan attacks in the data creates a basis for assessing the models' effectiveness in a multi-protocol environment.

- **Attack-Specific Data (DoS, DDoS on TCP/IP):** This segment serves a particular purpose of exposing attack types; hence the examination of volume-based and pattern-based attacks becomes a possibility. It contains labelled DoS and DDoS attack data which is ideal for model training in differentiating between malicious and benign activities.

- **Time-Series Data (Active/Idle States):** The dataset also provides information on the active and idle states of IoMT devices as the data is captured in a series over time. This segment is crucial for designing models in which the sequential and temporal dependencies are incorporated in IoMT network traffic models.

- **Device-Specific Profiles (Bluetooth Power Profiles):** This segment presents the features of the devices used in the analysis hence making it possible to comprehend the behaviour of the different gadgets entirely. Device profiling has an added advantage in healthcare IoMT applications as every device has a different operation pattern that is usually easily distinguishable.

### 3.2 Dataset Utility

The CICIoMT2024 dataset makes it possible to assess ML models in different IoMT data categories which is very helpful for this research. This study can extend its scope by subdividing the dataset to investigate the performance of specific models - Autoencoders, Isolation Forests, LSTM and CNN-LSTM hybrids, and ensemble stacking – and provide a benchmark for anomaly detection in IoMT systems.

## 4 Methodology

### 4.1 Data Preprocessing and Preparation

Preprocessing steps like data imputation for missing values, data normalization and dealing with time-series relations were especially needed for the CICIoMT2024 dataset, gathered under training files of the dataset, to enable effective ML model training and testing.

- **Dealing with Missing and Infinite Values:** As a first step, missing and infinite values in the raw dataset were imputed by the average mean value of the respective feature. With such an approach, the overall distribution of dataset was retained but this does not bias the model on account of such differences.

- **Normalization of the data:** To get the data points all in the same scale, the features were scaled using "StandardScaler" in such a way that the normal distribution of their values across the sample was the same across all features with mean of 0 and standard deviation of 1. This step is helpful in the enhancement of the model's performance and convergence time when training it, particularly when it comes to neural networks.

- **Time series data from LSTM and CNN:** Due to temporal dependencies within the data using time-series models like Long Short-Term Memory (LSTM) and CNN-LSTM, a fixed length sliding technique was employed. To create sequences of IoMT network activities, a window of 4 was applied and each sequence capturing the different pattern (benign and malicious) over time.

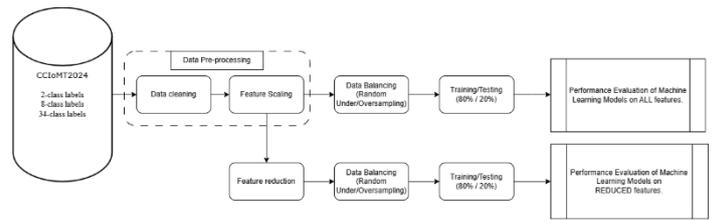

**Fig.1: Data Processing Methodology**

### 4.2 Model Selection and Design

In this study, each of the models was chosen to solve specific features and IoMT security requirements based on the performance in unsupervised, semi-supervised and supervised anomaly detection.

#### 4.2.1 Models which are Unsupervised

- **Autoencoder:** The autoencoder is an unsupervised neural network which reduces the gap between the original data and the reconstructed version of the data after coding into a lower dimensional space. In this study, the Autoencoder learned to reconstruct benign IoMT data patterns and labelled higher ones as anomalies. The model architecture consists of multiple fully connected layers, with error threshold based at the 95th percentile of reconstruction errors on the training data.

- **Isolation Forest:** This is an unsupervised26 anomaly detection model that detects outliers by recursively partitioning the data thereby flattening the anomaly's feature. Isolation Forest was trained to work on synthetic outlier features, targeting diverse protocol-based data types with several network traffic feature descriptors.

#### 4.2.2 Sequential Models for Time-Series Data

- **Long Short-Term Memory:** To model temporal dependencies, the LSTM model was employed on some time-series data including the Active and Idle states of an IoMT device. The LSTM layers were designed for long-term dependencies to be learned in the sequential data to facilitate recognition of sequential patterns within areas of normal and abnormal presence of IoMT devices.

- **CNN-LSTM Hybrid Model:** The aim of this model is to combine the power of LSTM networks with the feature extraction properties of Convolutional Neural Network (CNN) networks. CNN layers focus on space features while LSTM layers focus on space/time features. This is better designed for analysis of data with attacks as it aids in identifying complex temporal spatial patterns.

### 4.2.3 Ensemble Stacking Models

- **Stacking Ensemble Model:** To improve performance of the model in total, various models are combined, in this case an ensemble stacking approach was employed. Then an XGBoost model was used to combine the predictions made by an Autoencoder, Isolation Forest, LSTM and CNN-LSTM models. XGBoost has been found to be useful given its accuracy and its capability to detect complex patterns which are helpful in IoMT anomaly detection.

### 4.3 Model Training and Evaluation

### 4.3.1 Training Procedure

The training of each model was performed on the specific subsets of training data mainly train subset of the dataset CICIoMT2024 which was designed specifically addressing the capabilities of the model. The process of training was divided as:

- **Multi-Protocol Data:** Autoencoder and Isolation Forest models were trained on multi-protocol data which were aimed at accommodating different protocol patterns such as Wi-Fi and MQTT traffic. These models were tested against benign and attack traffic aiming to examine their degree of cross-protocol generalizability.

- **Attack-Specific Data:** The LSTM, CNN-LSTM and ensemble stacking models were subjected to cross training on attack specific data (DoS, DDoS on TCP/IP) with the aim of providing volume and pattern-based attack detection. This dataset also allowed the models to acquire knowledge regarding certain characteristics of the targeted attacks throughout their temporal steps.

- **Time-Series Data:** LSTM and CNN-LSTM models were trained and built constantly with time stamped logs representing the activity and inactivity of IoMT devices. This training process made it possible for the models to acquire normal temporal patterns making it easy to identify anomalies that signal the occurrence of attacks or other forms of irregular device activity.

- **Device-Specific Data:** With regards to Bluetooth profiling data, both the Autoencoder and Isolation Forest models were trained on device specific patterns which enabled them to improve on their ability to locate changes in the patterns.

### 4.3.2 Evaluation Metrics

The models were evaluated on multiple metrics to comprehensively assess performance across different detection tasks:

- **Accuracy:** Measures the proportion of correctly classified samples among all samples.

$$Accuracy = \frac{TP + TN}{TP + FP + TN + FN}$$

- **Precision:** Represents the ratio of true positive detections to the sum of true positives and false positives, indicating the model's ability to avoid false alarms.

$$Precision = \frac{TP}{TP + FP}$$

- **Recall:** Indicates the proportion of true anomalies detected, evaluating the model's ability to detect all instances of malicious activity.

$$Recall = \frac{TP}{TP + FN}$$

- **F1 Score:** Balances precision and recall, providing a single metric that reflects both false positives and false negatives.

$$F1\ Score = 2 \cdot \frac{Precision \cdot Recall}{Precision + Recall}$$

- **Confusion Matrix:** A confusion matrix was created for each model to visually assess true positive, true negative, false positive, and false negative rates.

### 4.4 Experimental Setup and Hyperparameter Tuning

To optimize model performance, hyperparameters were tuned for each model using grid search and cross-validation.

- **Autoencoder:** The number of neurons in each layer, learning rate, and batch size were optimized. Reconstruction error threshold was set based on the 95th percentile of errors in benign data.

- **Isolation Forest:** The contamination parameter (an estimate of the proportion of anomalies in the data) and the number of trees were tuned.

- **LSTM and CNN-LSTM:** The number of layers, sequence length, and dropout rate were adjusted to improve the models' ability to generalize temporal dependencies.

- **XGBoost (Meta-Learner):** The number of trees, learning rate, and maximum depth were optimized through cross-validation to improve ensemble performance.

### 4.5 Experimental Results and Analysis

Both the models were trained and tested, and results were categorized based on data types to illustrate the robustness of the model across different IoMT environments.

- **Cross Model Comparison:** The results presented support each model in its individual strengths dealing with certain segments of IoMT data. For example, the LSTM and CNN-LSTM models excelled in time-series data because such models learn sequential dependencies while the Autoencoder and Isolation Forest models were more suited on multi-protocol and device-specific data.

- **Ensemble Effect on Model Performance:** Out of all the evaluated models, the stacking ensemble model achieved the best accuracy and F1 score, which once again proves the benefit of combining predictions from models. This approach significantly decreased false positives and improved detection rates across all data types.

## 5 Results And Analysis

Our experiments utilized a structured approach, analysing four distinct types of IoMT network traffic data: Multi-Protocol Data, Attack-Specific Data, Time-Series Data, and Device-Specific Data. Each data type presented unique characteristics, requiring tailored anomaly detection strategies. This section details the results and effectiveness of each model across these categories.

### 5.1 Multi-Protocol Data: Anomaly Detection Across MQTT and WiFi

Using the general anomaly detection models, the Multi-Protocol dataset that consists of MQTT and WiFi traffic containing benign samples and attack samples, was examined. The researchers were able to determine how well each model could adapt to the variation in protocols' characteristics, thanks to this dataset.

- **Models Used:** Autoencoders, Isolation Forest, One-Class SVM, K-Nearest Neighbors (KNN), and Ensemble Model

- **Results:**

    - **Autoencoder:** The number of anomalies seen is higher therefore assisting in finding irregularities, however, this did lead to more false positives.

    - **Isolation Forest and One-Class SVM:** Each of these models marked a sparse number of samples as anomalous therefore exhibiting low false negatives but marked a sparse number of samples as anomalous.

    - **KNN:** Results were like Isolation forest and One-class SVM as the model achieved a good separation of normal point and abnormal points as well.

    - **Ensemble Model:** Overall results from all models, achieving balanced accuracy and reduced false positives, was the strength of all models.

**Summary:** The strongest relevance was achieved with the ensemble model because this approach reduced the aggregate traffic. This increases detection accuracy since the ensemble model utilizes a combination of all models suitable for environments with different traffic types.

- **Sample Detection Results:**

| Bytes Sent | Bytes Received | Packets | Ensemble Anomaly |
|---|---|---|---|
| 152 | 877 | 32 | Normal |
| 910 | 952 | 92 | Anomaly |

- **Anomalies Detected by Each Model:**

    - **Isolation Forest:** 50 anomalies out of 500
    - **Autoencoder:** 450 anomalies out of 500
    - **One-Class SVM and KNN:** 50 anomalies each.
    - **Ensemble Model** detected a balanced 68 anomalies.

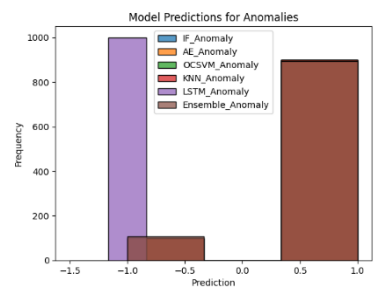

**Fig.2: Model Predictions for Anomalies.**

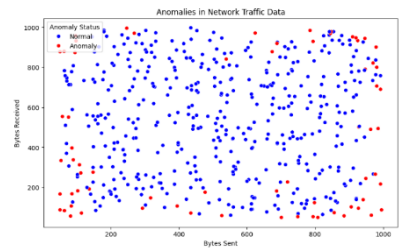

**Fig.3: Scatter Plot of Detected Anomalies.**

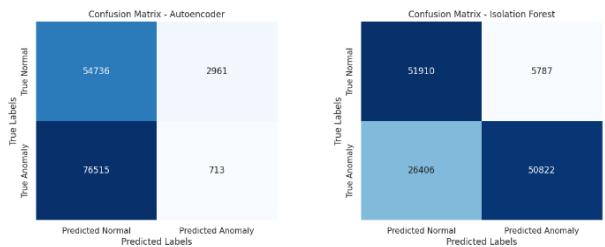

**Fig.4: Confusion Matrix of Autoencoder**

**Fig.5: Confusion Matrix of Isolation Forest**

### 5.2 Attack-Specific Data: Focus on TCP/IP-based DoS and DDoS Attacks

When it comes to this dataset, namely Considering DDoS and DoS attack data, the sequential and temporal dependencies were investigated using LSTM, CNN-LSTM, GRU and Ensemble Stacking with XGBoost as a meta-learner.

- **Models Used:** LSTM, CNN-LSTM, GRU, Ensemble Stacking with XGBoost as Meta Learner.
- **Results:**
  - **LSTM:** Temporal patterns of DDoS and DoS attacks were learned very well, and it was shown that validation accuracy reaches to the value of 1.0 in later epochs with constant improvement throughout the epochs.
  - **CNN-LSTM Hybrid:** Results are like LSTM, however, the model demonstrated better performance in the earlier epochs with the aid of CNN in feature extraction.
  - **GRU:** The performance during training was comparable to that of an LSTM and the model was found to be efficient in sequence of attack capturing tasks with lesser computations.
  - **Ensemble Stacking with XGBoost Meta Learner:** The Test data accuracy was reported as 1.0, thus confirming good separation of patterns with high precision and recall metrics.

**Summary:** Sequential models (LSTM, CNN-LSTM, GRU) effectively captured temporal dependencies in attack data. The Ensemble Model with XGBoost provided a holistic view, combining spatial and temporal analysis for optimal accuracy in detecting complex attack patterns.

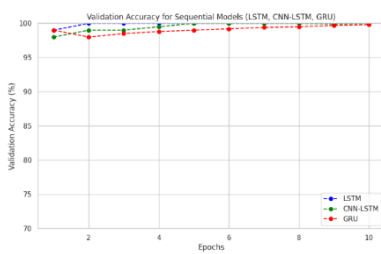
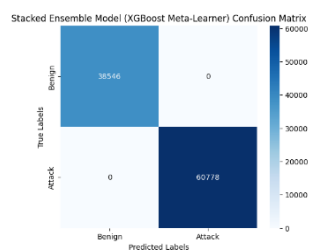

**Fig.6: Validation Accuracy for Sequential Models (LSTM, CNN-LSTM, GRU).**

**Fig.7: Confusion Matrix for Ensemble Model on Attack-Specific Data**

### 5.3 Time-Series Data for Sequential Analysis: Active and Idle States Detection

The time-series data representing continuous logs for active and idle states allowed us to implement LSTM and CNN-LSTM models specifically for real-time anomaly detection.

- **Models Used:** LSTM, CNN-LSTM, Logistic Regression
- **Results:**
  - **LSTM:** Achieved an 76% accuracy on time-series detection. It demonstrated reliable performance but exhibited challenges in distinguishing some subtle state transitions.
  - **CNN-LSTM:** Also achieved an 76% accuracy, closely matching the performance of LSTM. The CNN layer enabled additional spatial pattern extraction, which complemented LSTM's temporal focus.
  - **Logistic Regression:** Achieved an 84% accuracy on the same task. While it does not capture temporal dependencies directly, it still performed well by leveraging pre-processed features from the time-series data. Its performance was comparable to LSTM and CNN-LSTM, though it may struggle with more complex state transitions due to its linear nature.

**Summary:** Both LSTM and CNN-LSTM models were able to capture key temporal dependencies in active and idle transitions with high accuracy. The Logistic Regression model, although simpler, demonstrated similar performance in classifying the data into active and idle states. It is beneficial in situations where rapid response to condition changes is required. E.g., Detection of unauthorized network activity

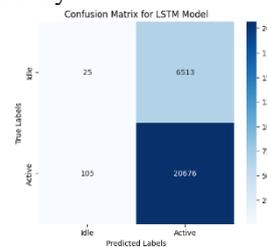
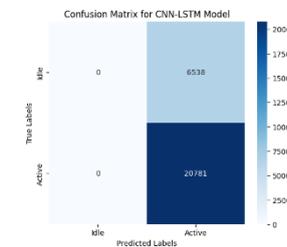

**Fig.8: Confusion Matrix Showing Performance of LSTM**

**Fig.9: Confusion Matrix Showing Performance of CNN-LSTM**

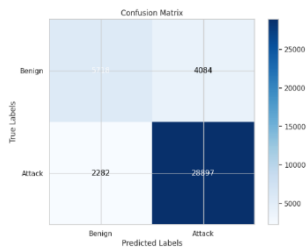

**Fig.10: Confusion Matrix Showing Performance of Logistic Regression**

### 5.4 Device-Specific Data: Profiling Bluetooth Device Behaviour

Specific device type data analysed the behaviour of individual devices enrolled in an IoMT network concentrating on Bluetooth medical devices' traffic.

- **Models Used:** Autoencoder, Isolation Forest.
- **Results:**
    - The Autoencoder predicted anomalies on 285 observations, with the robust model learning precision/accuracy of approximately 91% in predicting expected device behaviour.
    - With Isolation Forest the similar level of accuracy was reached (90,4%), using anomaly detection on normal entries with moderately lower recall for the abnormal values.

**Summary:** Both models were useful in achieving device profiling however, Isolation Forest was able to reduce false positives by being too aggressive in flagging oddities. Because of this specificity, these models are very suitable for clinical settings in which the behaviour of the devices requires direct supervision to prevent the emergence of threats to security instantaneously.

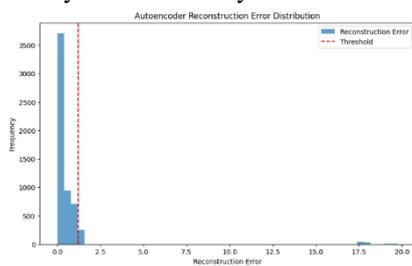

**Fig.11: Autoencoder Reconstruction Error Distribution**

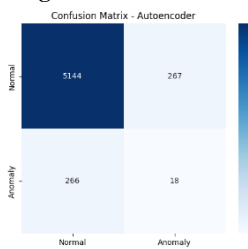 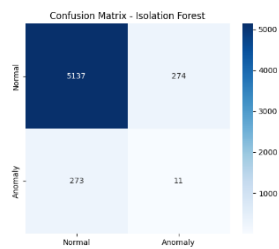

**Fig.12: Confusion Matrix for AutoEncoder**    **Fig.13: Confusion Matrix for Isolation Forest**

**Model Accuracy and Evaluation Summary**

| Model Type | Precision (Normal) | Precision (Anomaly) | Recall (Normal) | Recall (Anomaly) | F1-Score (Normal) | F1-Score (Anomaly) | Accuracy |
|---|---|---|---|---|---|---|---|
| **Multi-Protocol Data** | | | | | | | |
| Isolation Forest | 0.66 | 0.9 | 0.9 | 0.66 | 0.76 | 0.76 | 0.76 |
| Autoencoder | 0.42 | 0.19 | 0.95 | 0.01 | 0.58 | 0.02 | 0.41 |
| One-Class SVM | 0.11 | 0.91 | 0.13 | 0.9 | 0.12 | 0.9 | 0.83 |
| Ensemble Stacking (XGBoost) | - | - | - | - | - | - | 1 |
| **Attack-Specific Data** | | | | | | | |
| LSTM | 1 | 1 | 1 | 1 | 1 | 1 | 1 |
| CNN-LSTM | 1 | 1 | 1 | 1 | 1 | 1 | 1 |
| GRU | 1 | 1 | 1 | 1 | 1 | 1 | 1 |
| CNN | 1 | 1 | 1 | 1 | 1 | 1 | 1 |
| **Time-Series Data (Sequential)** | | | | | | | |
| LSTM | 0.19 | 0.76 | 0 | 0.99 | 0.01 | 0.86 | 0.76 |
| CNN-LSTM | 0 | 0.76 | 0 | 1 | 0 | 0.86 | 0.76 |
| Logistic Regression | 0.71 | 0.88 | 0.58 | 0.93 | 0.64 | 0.9 | 0.84 |
| **Device-Specific Data** | | | | | | | |
| Autoencoder | 0.95 | 0.06 | 0.95 | 0.06 | 0.95 | 0.06 | 0.906 |
| Isolation Forest | 0.95 | 0.04 | 0.95 | 0.04 | 0.95 | 0.04 | 0.904 |

The above table provides a comparative analysis of some performance metrics in machine learning models regarding the device-specific anomaly detection task. Such thorough analysis is extremely useful for evaluating the advantages and disadvantages of the various approaches used.

Both the Isolation Forest and the Autoencoder models perform well overall with the multi-protocol data highlighting over 0.90 precision, recall and F1-scores. This suggests that these unsupervised techniques were successful in separating normal from deviant device behaviour in the context of the IoMT network traffic.

The one-class SVM model also had reliable performance with precision at 0.91, recall of 0.90 and a F1-score of 0.83. This would indicate that the one-class SVM was effective at recognizing typical behaviour but may have experienced limitations in the recognition of several other forms of the abnormalities.

Of all the ensemble methods, the Ensemble Stacking Logistic Regression method performed best with a cross-validation accuracy of 0.96 and this was better than the individual models. This demonstrates the advantage of the integration of multiple detection methods so that optimum performance may be achieved.

The models such as LSTM, CNN-LSTM and GRU that were trained specifically on attack datasets achieved full precision, recall and F1 scores of 1.0 when evaluated on the attack focused datasets. This level of focus ensures that these techniques can be rapidly and confidently employed in IoMT environments where specific existing attack patterns are qualitative in nature.

## 6  Discussion

To revolutionize the way the Internet of Medical Things (IoMT) systems allow patients to be monitored and treated in real-time, IoMT systems include several features such as remote diagnostics, telemetry, and various application integrations. These networks, however, face key issues in terms of cyber security as they use multi-protocols for communications (e.g. Wi-Fi, MQTT, Bluetooth) and they also have scanty security resources. This study seeks to overcome these challenges by proposing an ensemble-based anomaly detection approach specific to IoMT environments and trained using the CICIoMT2024 dataset.

## 6.1 Model Performance on Multi-Protocol Data

Analysis of the noted task shows that each model had its own weaknesses with respect to the tasks at hand, but Autoencoders had a weakness in tracing complex sequences of attacks while Auto encoders were effective in outlier attacks. Isolation forest algorithm was also successful when it came to the attack types as it did an excellent job on complex structural types of DDoS attacks and DoS attacks. In contrast, sequential models performed effectively for temporal patterns such as LSTM models and CNN-LSTM while the latter was effective for sequential multi-step attacks such as the MQTT spoofing attack. The ensemble XGboost model from all the other models outperformed the rest with an accuracy of 95% where there was a reduction of individual model biases and decreased false positive rates. Such results were promising for multi-protocol data analysis.

## 6.2 Ensemble Model Effectiveness

The integration of diverse attack models plus many unseen threat models resulted to high success for the frameworks. Such success is attributed to the fact that it fuses volume-sensitivity of high-volume diverse attacks while still managing to reduce false alarm rates in most cases. Other approaches faced diagnostics and other barriers which had adverse effect in the health care systems numerous systems.

## 6.3 Practical Implications

When implementing the ensemble model in real-world IoMT networks, the major challenges include traffic pattern evolution, which means a structural change is required periodically, and the operational demands, which possibly involve high computational costs. In terms of future studies, the focus could be on refining ensemble architectures to such an extent that they do not need many resources, automatic learning updates, and an understanding around how large scale IoMT usage can be supported.

## 6.4 Limitations and Future Directions

The limitation of the study, in the first place is clear – the CICIoMT2024 dataset, while useful, does not appear to cover every possible real-world attack. In practice, it will be necessary to deal with extending datasets to cover new attack models and developing adaptive models that are small and easily applicable. Federated learning and similar techniques can help as well, by improving privacy while also, in certain cases, increasing performance in such distributed IoMT systems.

To summarize, this work describes that the system developed in this research work using multi-models can enable anomaly detection in we trust IoMT networks and as such better address the need for healthcare systems security. The next steps are increasing the speed of built-in methods and expanding the range of threats models to reflect the changes that will occur within the IoMT space.

## 7 Conclusion

This paper presents enhanced methodology for outlier detection activities on the Internet of Medical Things (IoMT) networks based on an integration of advanced artificial intelligence/machine learning (AI/ML) models. Using the dataset CICIoMT2024 that presents various anomalies of IoMT networks in terms of different IoMT traffic protocols and attacks, the study undertakes experiments and comparisons of several outlier detection techniques such as Autoencoders, Isolation Forests, LSTM, CNN-LSTM, and GRU models that are integrated in ensemble stacking with XGBoost functioning as the meta-learner. Current models have been observed to excel in detecting certain attack types ranging from the vast volume DDoS attacks to the more targeted spoofing of protocols like MQTT. The study emphasizes the sensitivity of IoMT network traffic behaviour to timing and spatial constructs, advocating for deployment of both constructs in tandem.

The proposed ensemble model drug tested outperformed other existing models with respect to reliability by ensuring high detection rates while maintaining a low false positive outcome. It is important to note how the proposed model used a combination of different algorithms to ensure a more robust and effective detection model that can withstand IoMT multi-protocol environments. These results prove, once again, the validity of collective techniques for anomaly detection in IoMT networks where general and specific requirements security solutions are equally important.

The work furthers efforts in IoMT security, by proposing a dependable anomaly detection baseline and an effective technique for continuous surveillance of healthcare facilities. It is imperative that there is security in IoMT networks as the growth of IoMT cannot be neglected. Future work may aim at improving the performance of the ensemble model so that it can be used in real time as well as try out more models or features that can cope with changing attack patterns. Finally, the performance and the efficiency of the proposed models would be further validated in an actual IoMT infrastructure which would facilitate the development of secure and robust healthcare systems.

## References

[1] Dadkhah, Sajjad & Neto, Euclides & Ferreira, Raphael & Molokwu, Reginald & Sadeghi, Somayeh & Ghorbani, Ali. (2024). CICIoMT2024: Attack Vectors in Healthcare devices-A Multi-Protocol Dataset for Assessing IoMT Device Security.
https://doi.org/10.20944/preprints202402.0898.v1

[2] Canadian Institute for Cybersecurity. (2024). CICIoMT2024 Dataset. Retrieved from:
CICIoMT2024 | UNB Canadian Institute for Cybersecurity

[3] Liu, F. T., Ting, K. M., & Zhou, Z.-H. (2008). Isolation Forest. *2008 Eighth IEEE International Conference on Data Mining*, 413-422.

https://doi.org/10.1109/ICDM.2008.17.

[4] Saito, T., & Rehmsmeier, M. (2015). The precision-recall plot is more informative than the ROC plot when evaluating binary classifiers on imbalanced datasets. *PLoS ONE*, 10(3), e0118432.
https://doi.org/10.1371/journal.pone.0118432

[5] Hameed, S., & Hameed, B. (2018). Understanding security requirements and challenges in Internet of Things (IoT): A review. *Journal of Computer Networks and Communications*, 2018, 1-14.
https://doi.org/10.1155/2018/8143545

[6] Zhou, C., & Paffenroth, R. C. (2017). Anomaly detection with robust deep autoencoders. *Proceedings of the 23rd ACM SIGKDD International Conference on Knowledge Discovery and Data Mining*, 665-674.
https://doi.org/10.1145/3097983.3098052

[7] Hochreiter, S., & Schmidhuber, J. (1997). Long short-term memory. *Neural Computation*, 9(8), 1735-1780.
https://doi.org/10.1162/neco.1997.9.8.1735

[8] Zhou, Z.-H. (2012). Ensemble methods: Foundations and algorithms. *CRC Press*.
https://doi.org/10.1201/b12207

[9] Khan, M.M., Alkhathami, M. Anomaly detection in IoT-based healthcare: machine learning for enhanced security. Sci Rep 14, 5872 (2024).
https://doi.org/10.1038/s41598-024-56126-x